\documentstyle[aps,aps12]{revtex}
\newcommand{\beq}{\begin{equation}}
\newcommand{\eeq}{\end{equation}}

\begin{document}

\sloppy
\parindent 6mm 
\parskip.13truein \baselineskip 22pt plus 2pt \vskip -8mm \hskip 80mm {\it %
slumin97.tex \ \ \ \ March 30, 1997} \vskip 30mm \centerline {\bf EVANESCENT
WAVES ARE NOT SUPERLUMINAL}

\vskip 25mm

\centerline {\ Michael Kelckner and Amiram Ron }

\centerline {\ Department of Physics }

\centerline {\ Technion - Israel Institute of Technology }

\centerline {\ Haifa 32000, Israel }

\vskip29mm

\centerline {\bf ABSTRACT}

\vskip10mm


It is demonstrated that an electromagnetic pulse, which is made to tunnel
through a barrier, would not be photo-detected before an identical pulse,
which travels the same distance in vacuum.


\vskip25mm 



\newpage


In the last few years several experiments were conducted on tunneling of
Electromagnetic (EM) signals through EM barriers~$^{1-4}$. In Refs. 1 and 2
microwave pulses were forced to overcome barriers as {\it evanescent} waves.
Steinberg~{\it et al}~$^{3}$ investigated the tunneling of a single-photon
across Multilayer Dielectric Mirrors (MDM). Speilmann~{\it et al}~$^{4}$
made 12 fs {\it classical} optical pulses to propagate through MDM. Both the
classical and the single-photon experiments appear to indicate {\it %
superluminal} electromagnetic tunneling.

In these experiments a single EM signal (a classical pulse or a
single-photon) was made to {\it split} into two equivalent signals. A
barrier was inserted along the path of one signal, and its "time of flight"
was compared with the other {\it reference} signal. The detection was
essentially a measurement of the {\it product} of the two signals: In the
classical case a nonlinear detector was employed, while in the quantum {\it %
twin photons} case the {\it overlap} of the wave packets was being detected.

In the present communication we readdress the issue of whether the EM
signal, which tunnel through the barrier, that is converted into an
evanescent wave, may be considered as moving {\it faster than light}. We
concentrate here on the classical case, where electromagnetic pulses can
clearly be represented by EM wave packets. We will only relate to the more
dramatic single-photon case by pointing out to the possible analogy between
a photon and a classical pulse. The important questions of different {\it %
velocities} like group velocity, signal velocity etc. are being
circumvented, by theoretically focusing on the direct detection of
photo-emission.


We start by considering the following experimental setup, which admittedly
may be only a {\it Gedanken} experiment. A classical electromagnetic pulse,
which propagates along the $x-axis$ with its electric field linearly
polarized in the $y-direction$ is split at the point $x=x_0$ into two
identical pulses. The two pulses are now propagating in parallel also along
the $x-axis$, with the above polarizations. Each of the pulses then travels
the same spatial distance $l$, where it reaches a photodetector at the point 
$x_l$. One of the pulses, which we label as the {\it reference} or "free"
pulse, travels all the distance $l$ in vacuum. While along the other path,
also in vacuum, of the second pulse, which we label as the {\it barrier}
pulse, a barrier of thickness $d$ is introduced, say between $x_1$ and $x_2$
with $x_2-x_1=d$ (See Fig.~1(a)). A barrier is understood here as a region
in space, where in the range of frequencies of the pulse, EM waves can
travel only as {\it evanescent} waves, and the transmitted wave is
attenuated without any absorption.

The EM signals will be represented by the electric field

${\bf E}({\bf r},t) = \hat{{\bf y}} E(x,t)$,

\noindent where ${\bf r}$ is the radius vector, and $t$ is the time. The
reference field propagating along the $x-axis$ is given by

\begin{equation}
E_r(x,t)= \int_0^{\infty} d \omega \ f(\omega - \omega_0) \ \cos \{ \omega [
t-t_0 - ( x - x_0 )/c ] \}. 
\end{equation}

\noindent Here $f(\omega - \omega_0)$ is the {\it spectral function} of the
pulse, $t_0$ is the time when the field is maximum at the {\it entrance}
position $x_0$, and $c$ is the speed of light in vacuum. The spectral
function of the angular frequency $\omega$ is taken to be symmetric and
centered at $\omega_0$. Its spectral {\it width}, $\Delta \omega$ is
extremely narrow, i.e. $\Delta \omega << \omega_0 $, however the {\it time
duration} of the pulse, $\sim 1/{\Delta \omega} $, is assumed to be much
shorter than $T = l/c $, the EM time {\it retardation} between $x_0$ and the
detector position at $x_l$. It is convenient to introduce the {\it analytic
signal},

$\overline E_r (x,t) = \int_0^{\infty} d \omega \ f(\omega - \omega_0) \
\exp \{ -i \omega [ t-t_0 - ( x - x_0 )/c ] \}$,

\noindent in terms of which $E_r(x,t) = \Re \{ \overline E_r (x,t) \} $,
where $\Re$ stands for the {\it real part of}.

The detector at $x_l$ is considered to be a {\it broadband} or {\it fast}
photoionization detector. The probability that the {\it first} electron will
be ejected by the incoming signal at the time $t$ is given by~$^5$

$P_{exc} (t) = \alpha \int_{t_0}^t dt^{\prime}\overline E_r^*
(x_l,t^{\prime}) \overline E_r (x_l,t^{\prime}). $

\noindent Here $\alpha$ is a parameter which characterizes the detector, and
the $^*$ denotes the {\it complex conjugate}. The lower limit of the time
integration is taken arbitrarily to be $t_0$, which is far in the past with
respect to the arrival time of the pulse to the detector. If we set, for
convenience, $t_0=0$, then for positive $t$ we get for the detector of the
reference signal

\begin{equation}
P_r (t) = \alpha \int_0^t dt^{\prime} \ \left \vert \int_0^{\infty} d \omega
\ f(\omega - \omega_0) \ \exp [ -i \omega (t^{\prime}-T) ] \right \vert^2. 
\end{equation}

We turn now to the detection of the barrier pulse. The analytic signal along
the {\it barrier} path, for $x > x_2$, i.e. to the right of the barrier is

$\overline E_b (x,t) = \int_0^{\infty} d \omega \ f(\omega - \omega_0) \
\tau( \omega) \ \exp \{ -i \omega [ t-t_0 - ( x - x_0 )/c ] \}$,

\noindent where $\tau(\omega)$ it {\it transmission} coefficient through the
barrier. While to the left of the barrier, for $x < x_1 $,

$\overline E_b (x,t) = \int_0^{\infty} d \omega \ f(\omega - \omega_0)$

$\ \ \ \ \ \left ( \ \exp \{ -i \omega [ t-t_0 - ( x - x_0 )/c ] \} + \rho(
\omega) \ \exp \{ -i \omega [ t-t_0 + ( x - x_0 )/c ] \} \right ) $,

\noindent   where $\rho(\omega) $ is the {\it reflection} function. For a
given physical barrier the functions $\tau$ and $\rho$ are determined
electromagnetically for a propagating {\it plane} wave of frequency $\omega$.

Thus for an identical broadband detector at $x_l$, on the barrier path, the
probability that the {\it first} electron will be ejected at time $t$ is
evidently

$P_{exc} (t) = \alpha \int_{t_0}^t dt^{\prime}\overline E_b^*
(x_l,t^{\prime}) \overline E_b (x_l,t^{\prime}) $.

\noindent In terms of the complex transmission function

$\tau(\omega) = B(\omega) \ \exp [ i \phi(\omega) ] $,

\noindent
where $B = \mid \tau(\omega) \mid $ and $\phi$ are real functions of the
frequency, we get for the detector of the barrier signal

\begin{equation}
P_b (t) = \alpha \int_0^t dt^{\prime} \ \left \vert \int_0^{\infty} d \omega
\ f(\omega - \omega_0) \ B(\omega) \ \exp [ -i \omega (t^{\prime}-T) +i
\phi(\omega) ] \right \vert^2. 
\end{equation}

In the present set-up the relevant question is which detector will fire
first, or rather - since the detectors must be treated quantum-mechanically
- which ionization probability is larger, $P_r(t)$ of the reference path, or 
$P_b(t)$ of the barrier channel. This way we avoid the questions related to
whether the velocity of the pulse is "faster than light". We shall
demonstrate that, at least for the {\it classical} experiments, when the
barrier prohibits {\it propagating} EM waves, i.e. allows only {\it %
evanescent} waves through it, the answer is $P_r(t) > P_b(t) $. That is, the 
{\it free} channel detector is more probable to fire first, and thus, in
this sense the evanescent wave is not superluminal.

We start with the analysis of an experimental set-up analog to that of
Ref.~4. It is clear that similar analysis would be relevant to the other 
{\it classical} cases of Refs.~1,2. We consider a barrier made of a {\it %
quarter wave stake} (aka MDM) of the form~$^6$ $(vacuum)(HL)^{11}(vacuum)$
fabricated of alternate layers of high refractive index material (titanium
dioxide, n=2.4), and low index material (fused silica, n=1.46). The pulse is
assumed to have a {\it Gaussian} spectral function $f$, centered at the
frequency $\nu _0\sim 375$ THz (wave-length $\lambda _0\sim 0.8\mu $m in
vacuum), and with a {\it bandwidth} $\Delta \nu \sim 28$ THz (FWHM). The
thickness of the alternate layers of the MDM is taken to be optically
equivalent to one quarter of the central wave length. The angle of incidence
of the {\it barrier} pulse on the MDM is taken to be zero. The complex {\it %
transmission coefficient}, $\tau (\omega )$, of this MDM is depicted in
Fig.~1(b). Here, the magnitude, $B(\omega )$ (solid line), and the phase, $%
\phi (\omega )$, (dashed line) are plotted as functions of the frequency;
and also shown, for comparison, is the spectral function of the pulse
(dash-dot line). It is evident that $B$ is {\it symmetric} with respect to
the central frequency, $\nu _0$, while $\phi $ is {\it antisymmetric}.
furthermore, the phase is practically {\it linear} over the range where $f$
is appreciable, and to a good extent its slope represents the apparent {\it %
time delay} seen in the experiments, e.g. $\sim 5$ fs of Ref.~4.

We now calculate numerically the relative probabilities for ejection of an
electron of the two {\it identical} detectors (same $\alpha $) using $%
P_r(t)/\alpha $ of Eq.~(2), and $P_b(t)/\alpha $ of Eq.~(3). The results are
plotted in Fig.~2, with the time measured relative to the vacuum retardation
time $T=(x_l-x_0)/c$. It is clearly demonstrated that, at {\it all times},
the probability of detecting an electron injected by the {\it vacuum} or
reference pulse, is much greater than that of the barrier pulse. The plateau
reached on the vacuum channel is about $10^4$ times larger than that of the
barrier channel, and is not seen in the figure. We therefore conclude that 
{\it direct} detection of the split pulses does not unveil any superluminal
behavior of the evanescent wave.

It would be interesting to verify this conclusion for a general barrier,
which compels propagating waves to {\it tunnel} as evanescent waves through
the barrier. It should be demonstrated that for any tunnel barrier, which is
represented by a causal $\tau(\omega)$, with attenuated transmitted pulse,
the difference in probability, $\Delta P(t) = P_r(t)-P_b(t)$, is positive.
From Eqs.~(2,3) we have

\begin{eqnarray}
   \Delta P (t) = & & \alpha \int_0^t dt'
     \int_0^{\infty} d \omega_1 \ f(\omega_1 - \omega_0)
     \ \exp [ i \omega_1 (t'-T)]
    \nonumber \\ & &
    \  \int_0^{\infty} d \omega_2 \ f(\omega_2 - \omega_0)
    \ \exp [-i \omega_2 (t'-T)]
       [ 1 - \tau^*(\omega_1) \tau(\omega_2) ].
   \end{eqnarray}                                     

\noindent We were not able to provide a general proof that $\Delta P(t) >0$,
however this seems to be plausible when in Eq.~(4) $\vert \tau(\omega) \vert
= B(\omega) $ is much smaller than one. As a matter of fact, this is indeed
the case, i.e. $B << 1 $, when the coincidence experiments~$^{1,2,4}$
apparently indicate "faster than light" results.

At this point we wish to indicate the difference between the previous
experiments, and the one which is analyzed here. Consider e.g. Spielmann 
{\it et al}~$^{4}$ experiment for comparison: The two channels for their
split pulse are the same as the reference pulse and the so-called barrier
pulse in our case. While we put {\it two} photodetectors at the end of the
two channels (of equal lengths), and compare the {\it direct} times of
flight, Spielmann {\it et al}~$^{4}$ use one nonlinear crystal as a
detector. They measure the intensity of the {\it Second Harmonic Generation}
(SHG) signal, due to time coincidence of the two pulses, which are directed
to the detector. Their control parameter is the arm length of the
reference pulse, which varies to produce coincidence. They detect
essentially the {\it 
the reference and barrier signals~$^{7}$, i.e.,

\begin{equation}
P_{SHG}(t)\propto \lim_{T \rightarrow \infty} 
{\int_0^T dt^{\prime} \left| \overline{E}_r(x_l,t^{\prime }+t )\right|
^2\left| \overline{E}_b(x_l,t^{\prime })\right| ^2}, 
\end{equation}

\noindent where $t$ is the relative delay time of the pulses. This
measurement of the {\it product} of the pulses is clearly sensitive to the
distortion, which the pulse suffers passing through the barrier. While in
our proposed scheme the response of the detectors is evidently connected to
the transfer times of the two pulses, in the coincidence experiments the
measurement corresponds to indirect evidence on the times of flight.


To conclude we wish to remark that the comparison with the case of Steinberg 
{\it et al}~$^3$ is harder to make, since they deal with single photons,
which should be analyzed by quantum mechanics. The only relevant statement
we may offer is that, if in some sense, single photons could have been
precisely described in terms of wave packets, analogous to the classical
ones outlined here, the same arguments would be applicable for a similar 
{\it Gedanken} experiment with two photodetectors for the twin photons.



{\it Acknowledgment.} This work was partially supported by the Franz
Ollendorff Chair and the Fund for Encouragement of Research in the Technion.



\newpage

{\bf References}

\begin{enumerate}
\item  A. Ranfagni, D. Mugnai, P. Fabini, G. P. Pazzi, G.Naletto, and C.
Sozzi, Physica (Amsterdam) {\bf 175B}, 283 (1991).

\item  A. Enders and G. Nimtz, J. Phys. I (France) {\bf 3}, 1089 (1993).

\item  A.M. Steinberg, P. G. Kwiat, and R. Y. Chiao, Phys. Rev. Lett. {\bf 71%
}, 708 (1993)

\item  Ch. Spielmann, R. Szipocs, A. Stingl, and F. Krausz, Phys. Rev. Lett. 
{\bf 73}, 2308 (1994)

\item  C. Cohen-Tannoudji, J. Dupont-Roc, and G. Grynberg, ''Atom Photon
Interactions'', Ch. II, p. 139, Wiley-Interscience, New York, (1992).

\item  M. Born and E. Wolf, ''Principles of Optics'', 6th edition, Pergamon,
NY, Section 1.6 (1980).

\item  R.G. Ulbrich and G.W. Fehrenbach, Phys. Rev. Lett. {\bf 43}, 963
(1979).
\end{enumerate}



\newpage

{\bf Figure captions}

\vskip 5mm

\noindent {\bf Fig. 1}

(a) A schematic description of the suggested setup. An electromagnetic pulse
is split at $x_0$ into two identical pulses. One pulse is propagating in
vacuum to $x_l$, where a photodetector (PD) is located. The second pulse is
made to tunnel a barrier along its path, and reaches a second detector, at
the same distance $l$. The photoionization probability of the detectors is
calculated as a function of time.



(b) The magnitude (solid line) and phase in units of $2\pi $ (dashed line)
of the transmission function through the Multilayer Dielectric Mirror are
plotted as a function of the frequency. Also shown for reference, a
schematic plot of the spectral function of the incident pulse ( in the
dash-dot line).

\vskip10mm

\noindent {\bf Fig. 2}

The calculated relative probabilities of photoionization of the two
detectors are plotted vs. time ( measured with respect to the vacuum
retardation time). 
It is seen that at all times the probability for photoionization of the {\it %
reference} detector is much larger than that of the {\it barrier} detector.






\end{document}